\documentclass[sigconf, authorversion]{acmart}

\usepackage{booktabs} % For formal tables
\usepackage{balance}

\acmSubmissionID{1162}

\copyrightyear{2018} 
\acmYear{2018} 
\setcopyright{acmlicensed}
\acmConference[ICMI '18]{2018 International Conference on Multimodal Interaction}{October 16--20, 2018}{Boulder, CO, USA}
\acmBooktitle{2018 International Conference on Multimodal Interaction (ICMI '18), October 16--20, 2018, Boulder, CO, USA}
\acmPrice{15.00}
\acmDOI{10.1145/3242969.3243014}
\acmISBN{978-1-4503-5692-3/18/10}

\begin{document}

\title{Attention-based Audio-Visual Fusion for Robust Automatic Speech Recognition}

\author{George Sterpu}
\authornote{Both authors contributed equally to this work.}
\affiliation{%
  \institution{ADAPT Centre, Sigmedia Lab, EE Engineering, Trinity College Dublin}
  \streetaddress{College Green}
  \city{Dublin}
  \state{Ireland}
}
\email{sterpug@tcd.ie}

\author{Christian Saam}
\authornotemark[1]
\affiliation{%
  \institution{ADAPT Centre, Sigmedia Lab, \\Trinity College Dublin}
  \streetaddress{College Green}
  \city{Dublin}
  \state{Ireland}
}
\email{saamc@scss.tcd.ie}

\author{Naomi Harte}
\affiliation{%
  \institution{ADAPT Centre, Sigmedia Lab, EE Engineering, Trinity College Dublin}
  \streetaddress{College Green}
  \city{Dublin}
  \state{Ireland}
}
\email{nharte@tcd.ie}

%\renewcommand{\shortauthors}{G. Sterpu et al.}

\begin{abstract}

Automatic speech recognition can potentially benefit from the lip motion patterns, complementing acoustic speech to improve the overall recognition performance, particularly in noise. In this paper we propose an audio-visual fusion strategy that goes beyond simple feature concatenation and learns to automatically align the two modalities, leading to enhanced representations which increase the recognition accuracy in both clean and noisy conditions. We test our strategy on the TCD-TIMIT and LRS2 datasets, designed for large vocabulary continuous speech recognition, applying three types of noise at different power ratios. We also exploit state of the art Sequence-to-Sequence architectures, showing that our method can be easily integrated. Results show relative improvements from 7\% up to 30\% on TCD-TIMIT over the acoustic modality alone, depending on the acoustic noise level. We anticipate that the fusion strategy can easily generalise to many other multimodal tasks which involve correlated modalities.
Code available online on GitHub: \url{https://github.com/georgesterpu/Sigmedia-AVSR}

\end{abstract}


% <concept_id>10010147.10010178.10010179.10010183</concept_id>
% <concept_desc>Computing methodologies~Speech recognition</concept_desc>
% <concept_id>10010147.10010257.10010293.10010294</concept_id>
% <concept_desc>Computing methodologies~Neural networks</concept_desc>
% <concept_id>10010147.10010178.10010224.10010225</concept_id>
% <concept_desc>Computing methodologies~Computer vision tasks</concept_desc>
% <concept_significance>300</concept_significance>
% <concept_id>10010147.10010257.10010258.10010259.10010265</concept_id>
% <concept_desc>Computing methodologies~Structured outputs</concept_desc>
% <concept_significance>300</concept_significance>
% <concept_id>10003120.10003121.10003124.10010870</concept_id>
% <concept_desc>Human-centered computing~Natural language interfaces</concept_desc>
% <concept_significance>300</concept_significance>

% \ccsdesc[500]{Computing methodologies~Speech recognition}
% \ccsdesc[500]{Computing methodologies~Neural networks}
% \ccsdesc[300]{Computing methodologies~Computer vision tasks}
% \ccsdesc[300]{Computing methodologies~Structured outputs}
% \ccsdesc[300]{Human-centered computing~Natural language interfaces}

\keywords{Lipreading; Audio-Visual Speech Recognition; Multimodal Fusion; Multimodal Interfaces}

\maketitle

\section{Introduction}

Human speech interaction is inherently multimodal in nature:  we both watch and listen when communicating with other people. Under clean acoustic conditions, the auditory modality carries most of the useful information, and recent state of art systems \cite{google_seq2seq} are capable of automatically transcribing spoken utterances with an accuracy above 95\%. The visual modality becomes most effective when the audio channel is corrupted by noise, as it allows us to recover some of the suppressed linguistic features. 

Exploiting both modalities in the context of Automatic Audio-Visual Speech Recognition (AVSR) has been a challenge. One reason is the inconclusive research on what are good visual features for Large Vocabulary Continuous Speech Recognition (LVCSR) \cite{review_pota} that match the well established Mel-frequency cepstral coefficients for acoustic speech. Another reason is the need for a fusion strategy of two correlated data streams running at different frame rates \cite{katsaggelos}.

Our work addresses both these challenges and we make a number of contributions. We introduce an audio-visual fusion strategy that learns to align the two modalities in the feature space. We embed this into a state of the art attention-based Sequence-to-Sequence (Seq2seq) Automatic Speech Recogition (ASR) system consisting of Recurrent Neural Network (RNN) encoders and decoders. We evaluate the strategy on a LVCSR task involving the two largest publicly available audio-visual datasets, TCD-TIMIT and LRS2, which contain complex sentences of both read speech and in-the-wild recordings. Using both of these datasets offers repeatability and allows other researchers to compare their systems directly to ours. We demonstrate improved  performance over an acoustic-only ASR system for clean speech and also for three  types of noise: white noise, cafe noise and street noise. 

The paper is organised as follows: In section \ref{sec:relwork} we review related work in the field of AVSR. In section \ref{sec:method} we describe our framework and introduce our audio-visual fusion strategy. Section \ref{sec:experiments} presents our experimental results, and we discuss our findings in Section \ref{sec:conclusion}.

\section{Related Work}
\label{sec:relwork}

\textbf{ASR.} The work of \cite{seq2seq_speech_comparison, google_seq2seq} shows that attention-based Seq2seq models \cite{attention_seq2seq} can surpass the performance of traditional ASR systems on challenging speech transcription tasks such as dictation and voice search. These models are characterised by a high degree of simplicity and generality, requiring minimal domain knowledge and making no strong assumptions about the modelled sequences. These properties make them good candidates for the AVSR task, where the potential benefit of the visual modality has been limited by an inability to exploit human domain knowledge for modelling. In our work, we use attention-based Seq2seq models to encode representations of both modalities.

\textbf{Lip-reading.} Transcribing speech from the visual modality alone, i.e. lip-reading, has improved significantly in recent years thanks to the advancements in neural networks. While there have been multiple contributions to the word classification task \cite{Chung16, petridis_classification, stafylakis, wand_improved}, the work of \cite{chung_cvpr_2017} transcribes large vocabulary speech at the character level, and this is the work most comparable to ours. The visual processing front-end consists of a Convolutional Neural Network (CNN) which learns a hierarchy of features from the training data. We adopt the CNN variant with residual connections \cite{resnet2}, as it attains competitive results in many computer vision tasks.

\textbf{Audio-Visual fusion.} Some of the earliest audio-visual fusion strategies are reviewed in \cite{review_pota, katsaggelos}, broadly classified into \emph{feature fusion} and \emph{decision fusion} methods. In AVSR, feature fusion is the dominant approach \cite{chung_cvpr_2017, petridis_classification}. Yet, as pointed out by \cite{review_pota}, concatenating features does not explicitly model the stream reliabilities. Furthermore, \cite{chung_cvpr_2017} observe that the system becomes over-reliant on the audio modality and apply a regularisation technique where one of the streams is randomly dropped out during training \cite{multimodal_2011}. Our method differs by always learning from the two streams at the same time. In addition, while \cite{chung_cvpr_2017} concatenates the sentence summaries, we learn to correlate the acoustic with the visual features at every time step, an opportunity to make corrections at much finer detail. Several other approaches from related areas focus on modelling cross-view or cross-modality temporal dynamics and alignment with modifications of gated memory cells \cite{ren_look_2016, rajagopalan_extending_2016, zadeh_multi-attention_2018, zadeh_memory_2018}. While these RNN architectures were introduced to address difficulties in learning long-term dependencies e.g. Long Short-Term Memory (LSTM) cells \cite{lstm} or Gated Recurrent Units \cite{D14-1179}, such dependencies were still found problematic in Seq2seq models based on gated RNN cells and lead to the introduction of temporal attention mechanisms \cite{attention_seq2seq}. Thus it seems reasonable to explore such attention mechanisms for potentially long-range cross-stream interactions rather than putting additional memorisation burdens on the basic building blocks.
A gating unit is introduced in \cite{tao_gating}, designed to filter out unreliable features. However, it still takes concatenated audio-visual inputs, and the synchrony of the streams is only ensured through identical sampling rate, limiting the temporal context. 

\section{Method}
\label{sec:method}

We begin by reviewing attention-based Seq2seq networks in order to introduce our fusion strategy more clearly in Section~\ref{subsec:avfusion}.

\subsection{Background}

The network consists of a sequence encoder, a sequence decoder and an attention mechanism. The encoder is based on an RNN which consumes a sequence of feature vectors, generating intermediate latent representations (coined as \emph{memory}) and a final state representing the sequence latent summary. The decoder is also an RNN, initialised from the sequence summary, which predicts the language units of interest (e.g. characters). Because the encoding process tends to be lossy on long input sequences, an attention mechanism was introduced to soft-select from the encoder memory a context vector to assist each decoding step.

\textbf{Attention mechanisms.} Attention typically consists in computing a context vector as a combination of state vectors (\emph{values)} taken from a memory, weighted by their correlation score with a target state (or \emph{query}).

\begin{eqnarray}
    \label{eq:attn1}
    \alpha_{ij} = score\ ({value}_j,\ {query}_i) \\
    \label{eq:attn2}
    c_i = \sum_j \alpha_{ij}\ {value}_j
\end{eqnarray}

The context vector \emph{c} is then mixed with the \emph{query} to form a context-aware state vector.
So far, this has been used successfully in Seq2seq decoders, where the \emph{query} is the current decoder state and the \emph{values} represent the encoder memory.

\subsection{Inputs}

Our system takes auditory and visual input concurrently. The \textbf{audio} is the raw waveform signal of an entire sentence. The \textbf{visual} input consists of video frame sequences, centred on the speaker's face, which correspond to the audio track. We use the OpenFace toolkit \cite{openface} to detect and align the faces, then we crop around the lip region.

\subsection{Input pre-processing}

\textbf{Audio input.} The audio waveforms are re-sampled at 22,050 Hz, with the option at this step to add several types of acoustic noise at different Signal to Noise Ratios (SNR). Similar to \cite{google_seq2seq}, we compute the log magnitude spectrogram of the input, choosing a window length of 25ms with 10ms stride and 1024 frequency bins for the Short-time Fourier Transform, and a frequency range from 80Hz to 11,025Hz with 30 bins for the mel scale warp. Finally, we append the first and second order derivatives of the log mel features, ending up with a feature of size 90 computed every 10 ms.

\textbf{Visual input.} The lip regions are 3-channel RGB images down-sampled to 36x36 pixels. A ResNet CNN \cite{resnet2} processes the images to produce a feature vector of \textbf{128 units} per frame. The details of the architecture are presented in Table~\ref{tab:resnet_details}.

\begin{table}[]
\centering
\caption{CNN Architecture. All convolutions use 3x3 kernels, except the final one. The Residual Block is taken from \cite{resnet2} in its \emph{full preactivation} variant.}
\label{tab:resnet_details}
\begin{tabular}{rcr}
%\hline
\textbf{layer} & \textbf{operation}                                      & \textbf{output shape} \\ \hline
0              & Rescale [-1 ... +1]                                                & 36x36x\textbf{3}               \\ %\hline
1              & Conv                                                    & 36x36x\textbf{8}               \\ %\hline
2-3            & Res block                                               & 36x36x\textbf{8}               \\ %\hline
4-5            & Res block                                               & 18x18x\textbf{16}              \\ %\hline
6-7            & Res block                                               & 9x9x\textbf{24}                \\ %\hline
8-9            & Res block                                               & 5x5x\textbf{32}                \\ %\hline
10             & Conv 5x5                                                & 1x1x\textbf{128}
\end{tabular}
\end{table}

\subsection{Sequence encoders}

Audio and visual feature sequences differ in length, being sampled at 100 and 30 Frames per Second (FPS) respectively. Across training examples, the sequences also have variable lengths. We process them using two LSTM \cite{lstm} RNNs. The architecture of the LSTMs consists of \textbf{3 layers} of \textbf{256 units}. We collect the top-layer output sequences of both LSTMs and also their final states, referring to them as encoding memories and sequence summaries respectively.

\subsection{Audio-Visual fusion strategy}
\label{subsec:avfusion}

This subsection describes the key architectural contribution. Our premise is that conventional dual-attention mechanisms, such as the one in \cite{chung_cvpr_2017}, overburden the decoder in Seq2seq architectures. In the uni-modal case, a typical decoder has to perform both language modelling and acoustic decoding. Adding another attention mechanism that attends to a second modality requires the decoder to also learn correlations among the input modalities.

\begin{figure}[t]
    \centering
    \includegraphics[width=0.95\linewidth]{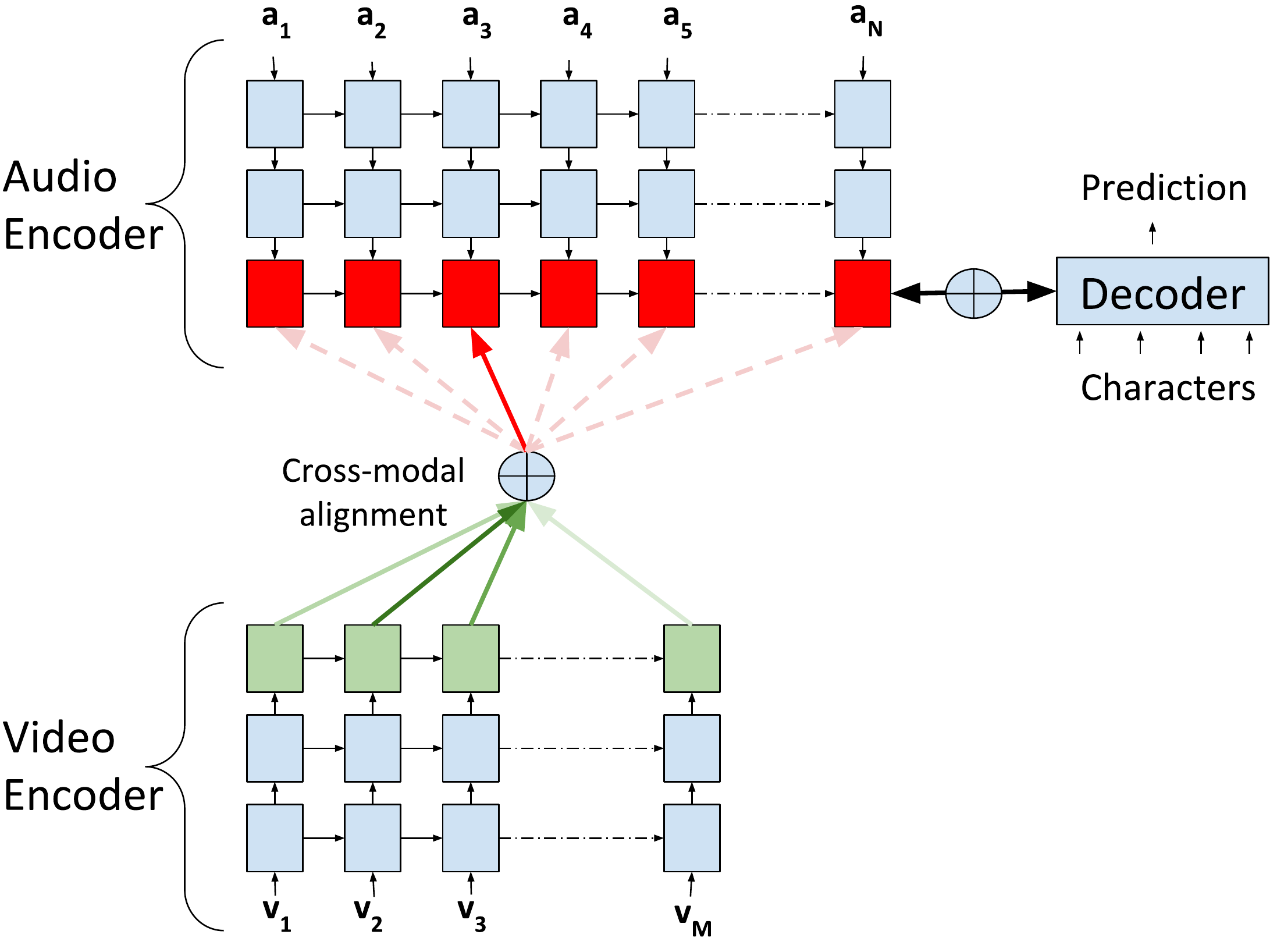}
    \caption{Proposed Audio-Visual Fusion Strategy. The top layer cells of the Audio Encoder (red) attend to the top layer outputs of the Video Encoder (green). The Decoder receives only the Audio Encoder outputs after fusion with the Video outputs. For clarity, the Decoder is not fully shown.}
    \label{fig:fusion}
\end{figure}

We aim to make the modelling of the audio-visual correlation more explicit, while completely separating it from the decoder. Thus, we move this task to the encoder side.
Our strategy is to decouple one modality from the decoder and to introduce a supplementary attention mechanism to the top layer of the coupled modality that attends to the encoding memory of the decoupled modality. The decoder only receives the final state and memory of the coupled encoder's top layer like a standard uni-modal attention decoder. A diagram of the fusion strategy is shown in Figure~\ref{fig:fusion}. 

The \emph{queries} from Eq.~(\ref{eq:attn1}) come from the states of the top audio encoder layer, while the \emph{values} represent the visual encoder memory.
The acoustic encoder's top layer can no longer be considered to hold acoustic-only representations. They are fused audio-visual representations based on corresponding high level features from the two modalities matched via attention. This layer could also be viewed separately as a higher level encoder (red layer in Figure~\ref{fig:fusion}) operating on acoustic and visual hidden representations.

The following intuition motivates our choice. The top layers of stacked RNNs encode higher order features, which are easier to correlate than lower levels. They provide speech related abstractions from visual and acoustic features. In addition, any time one feature stream is corrupted by noise its encodings may be automatically corrected by correlated encodings of the other stream.

The audio and visual modalities can have two-way interactions. However, by design only the acoustic modality learns from the visual. This is because in clean speech, the acoustic modality is dominant and sufficient for recognition, while the visual one presents intrinsic ambiguities: the same mouth shape can explain multiple sounds. The design assumes that acoustic encodings can be partially corrected or even reconstructed from visual encodings. One disadvantage is that an alignment score has to be computed for each timestep of the much longer audio sequence.

We also note that this fusion strategy explicitly models the alignment between each acoustic and all visual encodings. This elegantly addresses the problem of different frame rates, that traditionally required slower modalities to be interpolated in order to match the frame rate of the fastest modality.

\subsection{Decoding}

Our decoder is a single layer LSTM of \textbf{256 units}. As in \cite{google_seq2seq}, we use four attention heads to improve the overall performance, while still attending to a single enhanced memory. The decoder predicts characters, word level results are inferred by splitting at blanks.

\section{Experiments and Results}
\label{sec:experiments}

%\subsection{Task description}

We apply our audio-visual fusion method in the context of LVCSR. This is a well suited task for measuring the fusion performance, as correlations between modalities can be captured at the sub-word level. We also consider that the visual modality has a lot more to benefit from full sentences instead of isolated words, as longer contexts are critical to disambiguate the spoken message, thus supplying more informative features to the auditory modality. 

\subsection{Datasets}

TCD-TIMIT \cite{tcdtimit} and LRS2 \cite{mvlrs} are the largest publicly available audio-visual datasets suitable for continuous speech recognition. We report our results on both datasets.

\textbf{TCD-TIMIT} consists of 62 speakers and more than 6,000 examples of read speech, typically 4-5 seconds long, from both phonetically balanced and natural sentences. The video is laboratory recorded at 1080p resolution from two fixed viewpoints. We follow the speaker-dependent train/test protocol of \cite{tcdtimit}.

\textbf{LRS2} contains more than 45,000 spoken sentences from BBC television. Unlike TCD-TIMIT it contains more challenging head poses, but a much lower image resolution of 160x160 pixels.

Our results on LRS2 are not directly comparable with the ones on LRS \cite{chung_cvpr_2017}, which is completely different from LRS2 and was never publicly released. In addition, LRS2 has a lot more diversity in content and head poses than LRS. To the best of our knowledge, this is the first audio-visual baseline on LRS2.

We used the suggested evaluation protocol on LRS2 with one difference. The face detection \cite{openface} failed on several challenging videos, so we kept only those where the detection was successful for at least 90\% of the frames. With this rule, we kept approximately 87\% of the training videos and 91\% of the test videos. To foster reproducibility, we make available the list of files used in our subset.

% Finally, we apply text normalisation on the ground truth transcriptions, converting numbers from digits to characters.

\subsection{Training and evaluation procedures}

\begin{figure*}[h!]
    \centering
    \begin{minipage}{0.49\textwidth}
        \centering
        \includegraphics[width=1.0\linewidth]{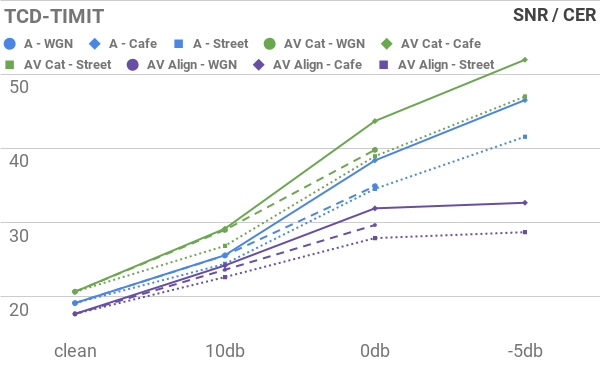}
        \caption{Character Error Rate  (CER) [\%] on TCD-TIMIT}
    \label{fig:tcd}
    \end{minipage}\hfill{}%
    \hfill
    \begin{minipage}{0.49\textwidth}
        \centering
        \includegraphics[width=1.0\linewidth]{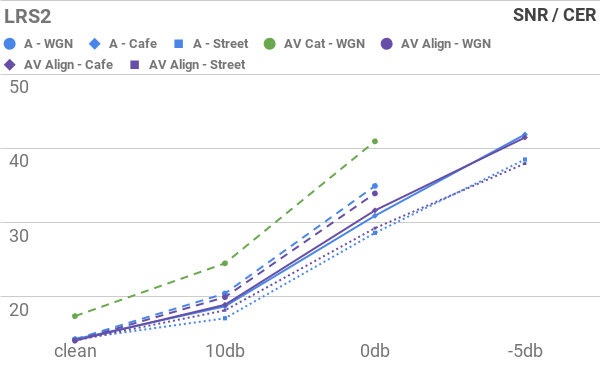}
        \caption{Character Error Rate  (CER) [\%] on LRS2}
        \label{fig:lrs2}
    \end{minipage}
\end{figure*}

We train several uni- and bimodal Seq2seq systems. 
The unimodal networks process only the acoustic input, either in clean form or additively corrupted by three types of noise: \emph{a) White Gaussian Noise (WGN)}, \emph{b) Cafeteria Noise}, and \emph{c) Street Noise}. The bimodal networks take both audio and visual inputs at the same time. We compare our method (\textbf{AV Align}) to an audio-only system (\textbf{A}) and dual attention feature concatenation \cite{chung_cvpr_2017} (\textbf{AV Cat}).

In training, we directly optimise the cross entropy loss between ground-truth character transcription and predicted character sequence via the AMSGrad optimiser \cite{amsgrad}. In evaluation, we measure the Levenshtein edit distance between them, normalised by the ground truth length. Our results report this Character Error Rate (CER) metric in percents. %We also split the  sequences at spaces and compute this metric at the word level, the Word Error Rate (WER). 
Our results are plotted in Figures~\ref{fig:tcd}-\ref{fig:lrs2}, while complete numerical values are listed in the appendix.  

\section{Discussion}
\label{sec:conclusion}

We begin by analysing our results on TCD-TIMIT. We first notice relative improvements starting at 7\% on clean speech (17.7\% CER down from 19.16\%), up to 30\% at -5db SNR (32.68\% CER down from 46.52\%) using our fusion strategy (\textbf{AV Align}) over audio only (\textbf{A}). The feature concatenation approach (\textbf{AV Cat}) performed between 8\% and 14\% worse than \textbf{A}, suggesting that the \textbf{AV Cat} strategy is far from optimal and may be difficult to train. Indeed, \cite{chung_cvpr_2017} reported training for 500,000 iterations. They also used a stream dropout regularisation technique which failed to improve convergence in our case after training for longer than \textbf{AV Align} took to converge.

Despite again clearly outperforming \textbf{AV Cat}, we see no benefit of \textbf{AV Align} over audio-only recognition (\textbf{A}) on LRS2. Our visual front-end, while effective on the limited variability conditions of TCD-TIMIT, may not be large enough to cope with the much more demanding conditions, varying poses and larger number of faces. Although not directly comparable, \cite{chung_cvpr_2017} used a more powerful front-end architecture with additional pre-training. It is plausible that an under-performing video feature extraction cannot provide strong enough support to the audio stream.

Further, we observe in the network outputs a progression through several stages of learning. At first, the decoder forms a strong language model learning correct words and phrases. Later, the influence of the acoustic decoding increases and the network learns letter to sound rules, over-generalising like a child and unlearning some of the correct spellings for words. The comparatively much larger size of LRS2 allows re-learning the multitude of letter to sound rules much more reliably, such that it may become the factor dominating the error rate and driving the training. We may then expect that the network needs to train longer and past this stage to start to fully exploit the visual information on LRS2.

With \textbf{AV Align} the network appears to learn how to cope with noise by leveraging the visual modality. Furthermore, few studies prior to \cite{chung_cvpr_2017} show improvements in the clean condition, prompting \cite{baltrusaitis_multimodal_2018} to call the interactions supplementary rather than complementary. That our model is able to consistently outperform audio only recognition on TCD-TIMIT even under clean conditions points to the inherent acoustic confusability of speech and the potentially sub-optimal speech recognition process for which the visual signal may offer some complementary information. We conclude that the \textbf{AV Align} strategy effectively separates the stream alignment task from the speech sound classification and language modelling tasks in the decoder, as opposed to the joint modelling in \cite{chung_cvpr_2017}. 

One shortcoming of our method is the lack of explicit modelling of the visual stream's reliability. However, the result on LRS2 suggests that the network may have learnt to discard noisy or uninformative visual features. 
An open question is whether the attention-based fusion strategy is powerful enough to cope with visual noise, rendering unnecessary a more complex structure. 

\section{Conclusions}

In this work, we introduce an audio-visual fusion strategy for speech recognition. The method uses an attention mechanism to automatically learn an alignment between acoustic and visual modalities, leading to an enhanced representation of speech. We demonstrate its effectiveness on the TCD-TIMIT and LRS2 datasets, observing relative improvements up to 30\% over an audio-only system on high quality images, and no significant degradation when the visual information becomes harder to exploit. Since our strategy was able to discover structure in an audio-visual speech recognition task, we expect it to generalise to others tasks where the input modalities are semantically correlated. 

Further to its performance, what makes our fusion strategy attractive is its straightforward formulation: it can be applied to attention-based Seq2seq approaches by reusing existing attention code. Not only may it effectively generalise to many other multimodal applications, but it also allows researchers to easily integrate our method into their solutions.

\begin{acks}
The \grantsponsor{1}{ADAPT Centre for Digital Content Technology}{https://www.adaptcentre.ie/} is funded under the SFI Research Centres Programme (Grant~\grantnum{1}{13/RC/2106}) and is co-funded under the European Regional Development Fund.
\end{acks}

\bibliographystyle{ACM-Reference-Format}
\balance
\bibliography{bibliography}

\clearpage
% \documentclass[sigconf]{acmart}

% \begin{document}

\appendix
\section{Numerical Results}

\begin{table}[h]
\centering
\footnotesize
\caption{Character Error Rate (CER) / Word Error Rate (WER)  [\%] on TCD-TIMIT}
\label{res:tcd}
\begin{tabular}{lllll}
% \hline

                      & \textbf{clean} & \textbf{10db} & \textbf{0db} & \textbf{-5db} \\ \hline

\textbf{A - WGN} & 19.16 / 45.53 & 25.58 / 53.89 & 34.92 / 64.98 & \\ %\hline
\textbf{A - Cafe} & 19.16 / 45.53 &  25.61 / 54.48 & 38.39 / 68.70 & 46.52 / 76.44 \\ %\hline
\textbf{A - Street} & 19.16 / 45.53 &  24.44 / 52.46 & 34.56 / 64.81 & 41.58 / 71.06 \\ %\hline

\hline

\textbf{AV Cat - WGN}   & 20.69 / 51.88 & 29.02 / 63.77 & 39.81 / 75.55 &  \\ %\hline
\textbf{AV Cat - Cafe} & 20.69 / 51.88 & 29.19 / 63.12 & 43.69 / 79/26 & 51.95 / 86.78  \\ %\hline
\textbf{AV Cat - Street} & 20.69 / 51.88 & 26.87 / 60.61 & 38.95 / 74.31 & 47.00 / 81.06  \\ %\hline

\hline

\textbf{AV Align - WGN}   & \textbf{17.70} / 41.90 & \textbf{23.65} / 49.95 & \textbf{29.68} / 57.07 &  \\ %\hline
\textbf{AV Align - Cafe}   & \textbf{17.70} / 41.90 & \textbf{24.23} / 51.94 & \textbf{31.93} / 60.88 & \textbf{32.68} / 60.58 \\ %\hline
\textbf{AV Align - Street}   & \textbf{17.70} / 41.90 & \textbf{22.66} / 49.73 & \textbf{27.92} / 55.19 & \textbf{28.72} / 55.63 \\ %\hline

\end{tabular}
\end{table}

\begin{table}[h]
\centering
\footnotesize
\caption{Character Error Rate (CER) / Word Error Rate (WER)  [\%] on LRS2}
\label{res:mvlrs}
\begin{tabular}{lllll}
%\hline
                      & \textbf{clean} & \textbf{10db} & \textbf{0db} & \textbf{-5db} \\ \hline
                      
\textbf{A - WGN} &  14.29  /  29.90          &    20.45 / 39.17        &    34.95   / 57.90      & \\ %\hline
\textbf{A - Cafe} & 14.29  /  29.95 &  \textbf{18.72} / 36.73 & \textbf{30.93} / 53.10  & 41.88 / 66.64 \\ %\hline
\textbf{A - Street } & 14.29  /  29.95 & \textbf{17.11} / 34.12 & \textbf{28.62} / 50.19 & 38.51 / 62.25 \\ %\hline

\hline

\textbf{AV Cat - WGN}   &     17.40  / 37.43         &     24.52 / 47.61         &  40.96  / 68.64    &   \\ %\hline

\hline

\textbf{AV Align - WGN}   &  \textbf{14.11} / 30.48 & \textbf{19.96} / 39.68 &  \textbf{33.94} / 57.89  &\\ %\hline
\textbf{AV Align - Cafe}   & \textbf{14.11} / 30.48 & 18.93 / 38.11 &  31.65 / 54.35  & \textbf{41.47} / 66.10 \\ %\hline
\textbf{AV Align - Street}   & \textbf{14.11} / 30.48 & 18.18 / 36.47 &  29.26 / 51.29  & \textbf{37.97} / 61.83\\ %\hline
\end{tabular}
\end{table}

% \end{document}

\end{document}